\documentclass[%
 reprint,
 amsmath,amssymb,
 aps,
]{revtex4-2}

\usepackage{lineno}
\usepackage{graphicx}
\usepackage{dcolumn}
\usepackage{bm}
\usepackage{float}
\usepackage{xcolor}

\begin{document}

\preprint{APS/123-QED}

\title{Blue Holes and Froude Horizons: \\
Circular Shallow Water Profiles for Astrophysical Analogs}

\author{Amilcare Porporato}
\email{aporpora@princeton.edu}
\affiliation{Department of Civil and Environmental Engineering and High Meadows Environmental Institute, Princeton University, Princeton, New Jersey 08540, USA}

\author{Luca Ridolfi}
\email{luca.ridolfi@polito.it}
\affiliation{Department of Environment, Land and Infrastructure Engineering, Politecnico di Torino, Corso Duca degli Abruzzi 24, 10129 Turin, Italy}
 
\author{Lamberto Rondoni}
\email{lamberto.rondoni@polito.it}
\affiliation{Dipartimento di Scienze Matematiche, Politecnico di Torino,
Corso Duca degli Abruzzi 24, 10129 Torino, Italy}
\affiliation{INFN, Sezione di Torino, Via P. Giuria 1, 10125 Torino, Italy}
 
\date{\today}

\begin{abstract}
Interesting analogies between shallow water dynamics and astrophysical phenomena have offered valuable insight from both the theoretical and experimental point of view. To help organize these efforts, here we analyze systematically the hydrodynamic properties of backwater profiles of the shallow water equations with 2D radial symmetry. In contrast to the more familiar 1D case typical of hydraulics, 
{even in} isentropic conditions, a solution with minimum-radius horizon for the flow emerges, similar to the black hole and white hole horizons, where the critical conditions of unitary Froude number provide a unidirectional barrier for surface waves. Beyond these time-reversible solutions, a greater variety of cases arises, when allowing for dissipation by turbulent friction and shock waves (i.e., hydraulic jumps) for both convergent and divergent flows. 
The resulting taxonomy of the base-flow cases may serve as a starting point for a more systematic analysis of higher-order effects 
linked, e.g., to wave propagation and instabilities, capillarity, variable bed slope, and rotation.  \end{abstract}

\maketitle

\begin{widetext}
\begin{flushright}

{\it “(l’Idraulica), sempre da me tenuta per difficilissima e piena di oscurità.”

Galileo Galilei [letter to R. Staccoli, January 16, 1630]}
\end{flushright}
\end{widetext}
\section{Introduction}

Browsing the recent literature on fluid analogies in astrophysics \cite{unruh1981experimental,Barcelo2005}, Galileo may object that his quote about the 'obscurities of hydraulics' actually does not refer to the darkness of the black holes, but to the mysteries of fluid dynamics, which still persist today. Indeed, it was not until the early 1970's, that striking analogies between the properties of black holes and thermodynamics were pointed out, associating the area of a black hole event horizon with the thermodynamic entropy. A kind of second law was proven to hold \cite{Beke1972,Beke1973,Hawk1973,Hawk1975} (cf.\ Ref. \cite{Page2005} for a review) and a correspondence with hydrodynamic systems was observed \cite{Smarr1973}. 
Whether such analogies are to be taken as expressions of {\em bona fide} thermodynamic properties was an issue at the time, and so remains today \cite{Wallace2010,Wallace2018,Call2020}.

In 1981, Unruh \cite{unruh1981experimental} noted that there exist well understood acoustic phenomena, reproducible in the laboratory, which formally enjoy the same properties of black holes, as far as the quantum thermal radiation is concerned (the acoustic metric is conformal, but not identical, to the Painlev\'e–Gullstrand form of the Schwarzschild geometry \cite{Barcelo2005}). { Independently, acoustic analogies were considered also by other authors, notably by Moncrief \cite{Moncrief1974acoustic} and Matarrese \cite{Matarrese1985perfectfluid}. In particular, Unruh's work}
implies that exotic phenomena, such as black hole evaporation, may be tested in controllable experiments on Earth! The way was paved for a research line now known as {\em analogue gravity} \cite{Barcelo2005}, which has led to numerous experimental analogues of cosmological gravitational phenomena. 

Most notable for their 
beauty and variety are the hydraulic experiments, which in turn are connected by analogy to polytropic gasdynamics  \cite{landau1987theoretical,whitham2011linear}. For instance, Ref. \cite{foglizzo2012shallow} reports on a shallow water experiment mimicking the development of a shock instability in the collapse of a stellar core, { that makes highly non-spherical the birth of a neutron star, in spite of} the underlying spherical symmetry. The element of interest here 
is the 2-dimensional hydraulic jump, in a convergent radial flow, which corresponds to the accretion spherical shock that 
arises above the surface of the neutron star when it is being generated. The diverging radial case is instead associated to  
a white hole, a kind of {\em time reversed} black hole \cite{Jannes2011,Bhattacharjee2016}, which is experimentally 
realized even with superfluids \cite{Volovik2005,Volovik2006}. In Ref. \cite{Euve2020}, the scattering of surface waves 
on an accelerating transcritical 1-dimensional flow are investigated, as corresponding to the space time 
of a black hole horizon, while in Ref. \cite{schutzhold2002gravity}, gravity waves have been used to investigate instabilities 
of both black and white holes. These are but a few examples (see also \cite{rousseaux2008observation,weinfurtner2011measurement,euve2016observation}) out of the many one can find in the specialized literature.

Using the same mathematical framework to describe very diverse physical situations has repeatedly proven 
successful. Laplace considered Newtonian mechanics equally suitable to describe the {\em 'movements of the greatest bodies 
of the universe and those of the tiniest atom'} \cite{Laplace1829}. Classical mechanics is effective not only when dealing with several macroscopic objects, under non-relativistic conditions, but 
it also constitutes the basis of statistical mechanics, { which deals} with huge numbers of atoms and molecules. The planets orbiting the Sun and the molecules in a glass of water are separated by 52 orders of magnitude in mass, and by 23 orders of magnitude in numbers. The shallow water experiment and the collapse leading to a neutron star \cite{foglizzo2012shallow} are separated, after all, by  only six orders of magnitude { in size.}

Within this context, our contribution aims at providing a systematic classification of the so-called backwater profiles (i.e., the elevation of the free surface as a function of the streamwise coordinate), possibly connected by shock waves (i.e., the hydraulic jumps), as solutions of the shallow water equations in circular symmetry. A simple but systematic discussion of this type may have the merit of connecting apparently different flow configurations. While most of the shallow water literature refers to 1D streams, here we focus on the role of the circular symmetry enforced by the continuity equation, which opens the possibility for convergent or divergent flows; we also pay attention to the role of dissipation, provided not only by friction but also, when present, by shock waves in the form of hydraulic jumps.

The astrophysics literature has mainly dealt with cases with circular hydraulic jumps, either convergent \cite{foglizzo2012shallow} or divergent \cite{Jannes2011,Bhattacharjee2016}, as well as with 1D currents with critical transitions over obstacles \cite{rousseaux2008observation,weinfurtner2011measurement}. While these dissipative cases are characterized by strong energy losses due to the presence of hydraulic jumps, and therefore are not symmetric with respect to time or velocity inversion, here we also emphasize the presence of an inviscid solution, which obeys this symmetry. 
The corresponding analytical solution is of particular interest, as it represents a close analog of the black hole; herein, the subcritical (i.e.,\ subluminal, in the analogy) convergent current accelerates towards a 'Froude horizon', where the velocity becomes critical (i.e.,\ equal to the speed of the surface waves). The white hole analogy emerges  naturally, when the flow velocity is reversed. 

Turbulent friction, acting in the direction opposite to the flow, allows for the appearance of a critical point in the dynamical system describing the free surface profiles, and thus { it} indirectly allows for the presence of stable hydraulic jumps, represented as sharp discontinuities in the surface profiles. Depending on the boundary conditions, both converging and diverging cases with no jumps (called here dissipative black and white holes, respectively) are also found, along with the corresponding cases with shock waves.

We confine our discussion to steady state profiles, in both inviscid and turbulent conditions, noticing however that the laminar case presents no qualitative differences. While the interesting effects of capillarity and rotation are left for future work, we hope that the present analysis may be useful to provide a classification of base flows to analyze systematically the links between shallow water profiles and their astrophysical analogs.

\section{Governing Equations}

The starting point of the analogy are the well-known shallow water equations \cite{landau1987theoretical,whitham2011linear}, namely the continuity equation
\begin{equation}
\label{eq:cont}
\partial_t h= - \boldsymbol{\nabla}\cdot(h \boldsymbol{v}) \, ,
\end{equation}
where $h$ is the water depth, and $\boldsymbol{v}$ is the depth-averaged velocity, { which obeys} the momentum equation
\begin{equation}
\label{eq:mom}
\partial_t \boldsymbol{v}+\boldsymbol{v}\cdot \boldsymbol{\nabla}\boldsymbol{v}+g\boldsymbol{\nabla}h=\boldsymbol{\nabla}z_b-\boldsymbol{j},
\end{equation}
where $g$ is the gravitational acceleration, considered constant, $z_b$ is the bed elevation, and $\boldsymbol{j}$ is the frictional force. 

Friction is modeled as
\begin{equation}
\label{eq:fric}
\boldsymbol{j}=\frac{|\boldsymbol{v}|^\alpha}{C^2 h^\beta}\boldsymbol{v},
\end{equation}
where $C^2$ is the Chezy's coefficient, which is inversely proportional to the friction coefficient. 

In what follows, we will only consider horizontal stream beds, $\boldsymbol{\nabla}z_b=0$, and will focus on the case of $\alpha=1$ and $\beta=1$, typical of fully developed turbulent flows. For simplicity, we assume $C^2$ to be constant, according to Bresse's hypothesis \cite{bresse1860cours}. More complicated formulations \cite{bonetti2017manning,bohr1997averaging,luchini2010phase,ivanova2019structure}, including the laminar case with $\alpha=0$, $\beta=2$ \cite{bohr1993shallow}), do not change the picture qualitatively. Other effects, such as rotation and capillarity, although potentially very interesting, are neglected here. We will return to a discussion of their effects and to possible extensions of this work in the concluding section.

\section{Isentropic case}

\begin{figure*}
  \centering
  \includegraphics[width=150mm]{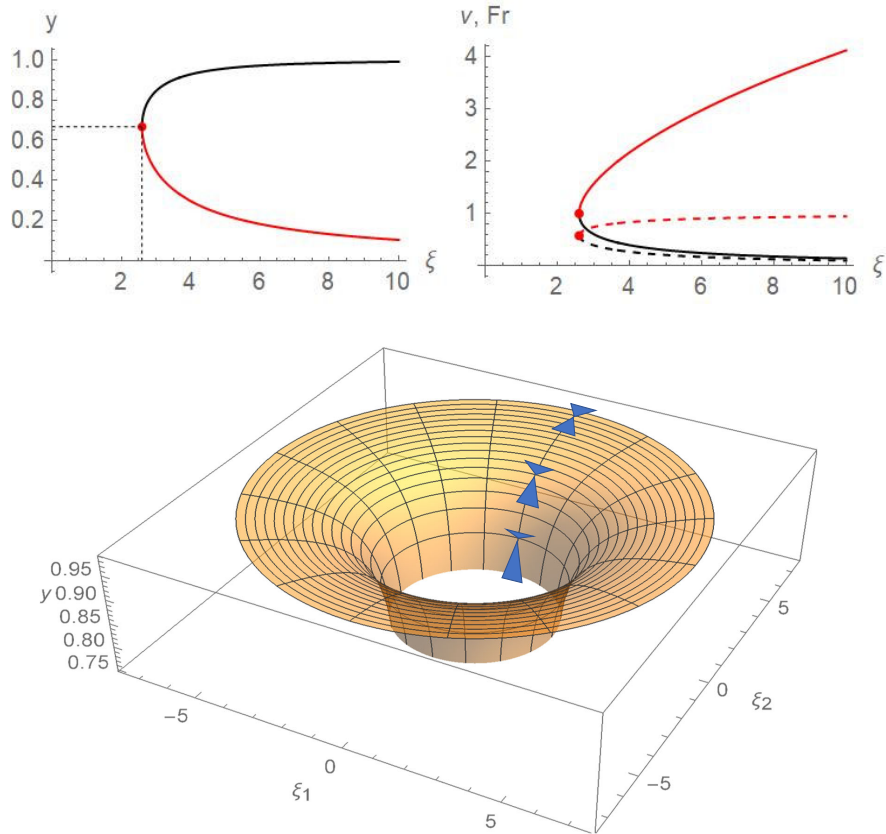}
    \caption{Isentropic (inviscid) case. Top left panel: stream profiles as given by Eq.  (\ref{eq:xi}); the red point marks the critical condition $\xi_{min}={3\sqrt{3}}/{2}$, $y(\xi_{min})={2}/{3}$. Top right panel: behavior of the  velocity (dashed lines) and Froude number (solid lines) in the subcritical (black lines) and supercritical (red lines) conditions. Bottom panel: 3D view of the dimensionless water-level profile for the subcritical branch of the inviscid solution (\ref{eq:xi}), corresponding to the black hole. A qualitative rendering of the 'light cones' for the propagation of small waves is also shown.}
  \label{fig:profile_inviscid}
\end{figure*}

When reduced to a circularly symmetric problem, in steady state and in the absence of friction, the continuity equation becomes 
\begin{equation}
\label{eq:cont1}
2 \pi r v h= Q,
\end{equation}
where $Q$ is the total volumetric flowrate, while the 
momentum equation { takes the form} 
\begin{equation}
\label{eq:mom2D}
\frac{d}{dr}\left(h+\frac{v^2}{2g} \right)=0,\end{equation}
{ and it} can be integrated as
\begin{equation}
\label{eq:mom2Dinv}
h+\frac{v^2}{2g}= H,\end{equation}
where $H$ is the constant head (energy per unit weight of fluid). 

Eliminating the velocity from the previous equations, 
\begin{equation}
\label{eq:mom2Dinv1}
h+\frac{Q^2}{2g (2 \pi r)^2 h^2}= H,\end{equation}
consistently with \cite{watson1964radial,hager1985hydraulic,bohr1993shallow}, suggests a normalization with $y=h/H$ and $\xi=\frac{r}{R}$, where $R=\frac{Q}{2 \pi H \sqrt{2gH}}$ is the radius at which the given discharge passes with Torricellian velocity $\sqrt{2gH}$ and height $H$. As a result, 
\begin{equation}
\label{eq:xi}
\xi=\frac{1}{y \sqrt{1-y}}.\end{equation}
where, by definition, one has $\xi > 0$ and $y \in (0,1)$.
The solutions are reported in Fig. 1, along with the corresponding velocity, obtained using
\begin{equation}
\nu=\frac{1}{\xi y}.   
\end{equation}
Note that $\xi_{min}=\frac{3\sqrt{3}}{2}$ and $y(\xi=\xi_{min})=\frac{2}{3}$ (see Eq. (40) in \cite{schutzhold2002gravity}). Each branch of the solution represents two possibilities, since the flow direction can be inverted, because the equations are invariant 
under changes in the flow direction. In the astrophysical analogy, the top branch represents a black hole, when the flow is convergent, and a white hole, when the flow is divergent, both with no dissipation.

The occurrence of a minimum radius, $\xi_{min}$, can be understood by considering the radian-specific discharge, $Q_u=Q/2 \pi r$, or $q_u=1/\xi$ in dimensionless form, which combined with the stream profile (\ref{eq:xi}) yields 
\begin{equation}
\label{eq:qu}
q_u=y\sqrt{1-y}\end{equation}
showing that $q_u$=$q_u(y)$ is a non-monotonic relation, with a maximum at $y=2/3$, 
corresponding to $q_u=1/\xi_{min}$. 
This limit behavior arises because 
a stream approaching smaller radii (flowing according to one of the two branches) can 
carry that flow per radian only until the maximum value of $q_u$ is attained, and the 
stream reaches the depth $y=2/3$ at $\xi_{min}$. For smaller radii, the stream cannot carry that much energy $H$ and discharge $Q$, while conserving them. The 
condition of maximum $q_u$ is called {\em critical} (i.e., $y=2/3 \equiv y_c$). 

By introducing the Froude number
\begin{equation}
\label{eq:Fr}
Fr=\frac{v}{\sqrt{gh}}=\frac{1}{\xi} \sqrt{\frac{2}{y^3}}, \end{equation}
the critical condition corresponds to $Fr=1$, { which is plotted as { the} red line $y_c=(\sqrt{2}/\xi)^\frac{2}{3}$ in Fig. \ref{fig:profile_inviscid}. Accordingly,} the upper (lower) branches showed in Fig. \ref{fig:profile_inviscid} are called subcritical (supercritical) conditions. Notice that the  Froude number is the ratio between the stream velocity, $v$, and the propagation celerity of small-amplitude surface waves, $\sqrt{gh}$ (first obtained by Lagrange \cite{Lagrange}). It follows that subcritical streams are characterized by surface waves that can propagate against the stream ($Fr<1$); in contrast, waves can only propagate in favor of current if streams are supercritical ($Fr>1$). This is also reflected in the 'light cones' drawn in Fig. 1 (lower panel).

The interplay between the stream and the wave velocities and the existence of a minimum radius are analogous to the conditions of the Schwarzschild solution of the black hole. Like the latter corresponds to a mass confined in a region, whose escape velocity equals that of light, the minimum radius $\xi_{min}$ corresponds to the threshold at which surface waves are no longer able to go up the current. As a subcritical current flowing towards the central hole with decreasing water depth $h$, the stream velocity $v$ increases, while the wave celerity $\sqrt{gh}$ decreases. It follows that the Froude number gradually approaches 1, and the stream reaches this critical value precisely at a 'Froude horizon', 
inside which the surface waves are no longer able to go upstream, namely to run away from the hole. In this sense, the surface waves resemble the light in the Schwarzschild problem. However, while in the black hole the velocity of the falling observer decreases proportionally to the square root of the radius and the light speed remains constant, here (see top-right panel in Fig. 1) the surface-wave speed changes in space depending on the water depth; as mentioned in the Introduction, 
the correspondence between the black hole and the shallow water metrics is only conformal \cite{Barcelo2005}. 

Note that modifying the 
dependence of the bottom slope with the radius, the analogy could possibly be made tighter, as 
also briefly noted in \cite{schutzhold2002gravity} 
but this will not be pursued here. Another interesting point for future work regards the conditions beyond the minimum radius, namely inside the Froude horizon, where different solutions, similar to the so-called interior Schwarzschild solution \cite{hughston1990introduction}, might 
account for the fact that the flow cannot take place with the same discharge and energy.  

\section{Backwater profiles due to friction}

Allowing for the effects of turbulent friction, several additional configurations become possible. These can be obtained considering, in terms of water depth, the combined continuity and momentum equation as
\begin{equation}
\label{eq:mom2D1}
\frac{dH}{dr}=\frac{d}{dr}\left(h+\frac{v^2}{2g} \right)=-\frac{v|v|}{C^2 h}.\end{equation}
Using the normalization $v/\sqrt{2g H_0}=\nu=1/(\xi y)$ (where $H_0$ is the stream head at { the boundary)} 
and the other reference scales introduced before yields
\begin{equation}
\label{eq:mom2D2}
\frac{dy}{d\xi}\left(1-\frac{2}{\xi^2 y^3} \right)=\frac{2}{\xi^3 y^2}-\frac{\alpha}{\xi^2 y^3},\end{equation}
where $\alpha={\rm sgn}(v)2g/C^2$. Thus $\alpha$ is positive for turbulent flows proceeding along $\xi$ (divergent) and negative for flows that go against $\xi$ (convergent). As a result, the slope of the water depth is
\begin{equation}
\label{eq:mom2D4}
\frac{dy}{d\xi}=\frac{2y-\alpha\xi}{\xi^3 y^3-2\xi}=\frac{N(\xi, y, \alpha)}{D(\xi,y)}.\end{equation}

\begin{figure*}
  \centering
  \includegraphics[width=150mm]{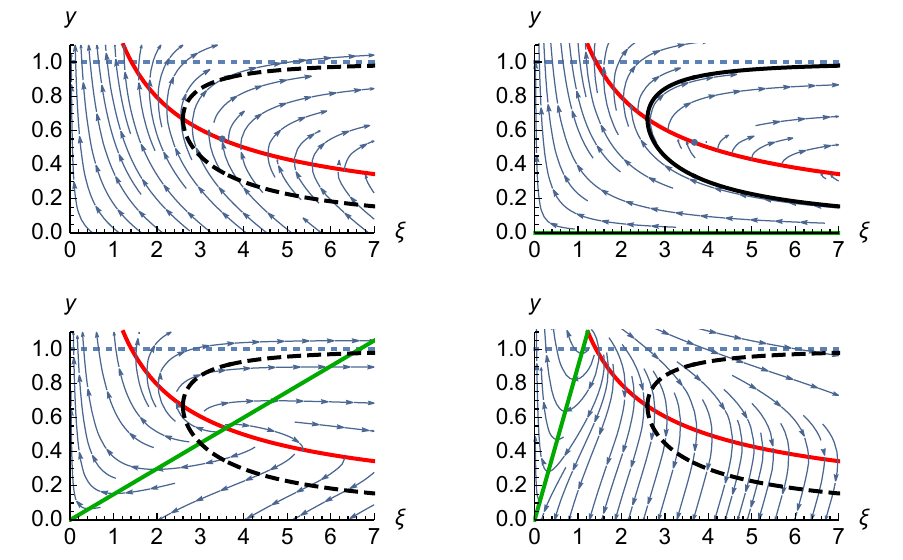}
  \caption{Phase space for $\alpha$ increasing in steps of 0.4, starting from -0.4 (top left). The red line is $D=0$, while the green is $N=0$; the black line is the solution of the inviscid case, Eq (\ref{eq:xi}).}
  \label{fig:phase}
\end{figure*}

Figure \ref{fig:phase} shows the phase plots for some values of $\alpha$. In these plots, three curves are highlighted: the profiles (\ref{eq:xi}) corresponding to the inviscid case (black lines), and the solutions of equations $N(\xi, y, \alpha)=0$ (green lines) and $D(\xi, y)=0$ (red lines). Since the physical domain is bound to $\xi>0$, green lines occur only if $\alpha>0$.

In the $\alpha=0$ case (i.e., no energy dissipation), the inviscid solution reproduces the only possible stream profiles compatible with energy conservation: depending on the boundary condition, the subcritical or the supercritical reach is selected (notice that the critical condition, where the reaches join, lies on the curve $D(\xi, y)=0$).
Differently, when dissipation occurs (i.e., $\alpha \ne $0), the curve representing the flow starts from the boundary 
with energy $H_0$ and flows dissipating energy according to the Eq. (\ref{eq:mom2D1}). Since $H_0$ was chosen as a reference energy to normalize the problem, it follows that the boundary condition lies on the inviscid solution, where $H(\xi_{bc})/H_0=1)$, being $\xi_{bc}$ the radial position of the  boundary. The remaining part of the stream profile follows the curve of the phase space, departing from the boundary condition. In other words, the black curve corresponding to the solution (\ref{eq:xi}) now becomes the locus of the initial conditions of the backwater profiles. As expected, the phase trajectories are tangent to the inviscid solution only for $\alpha=0$.   

Becuase of dissipation, the flows are characterized by $(h+v^2/2g) \leq H_0$, that in dimensionless form reads
\begin{equation}
\label{eq:en_constr}
y+\frac{1}{\xi^2 y^2} \leq 1.
\end{equation}
Such a condition is satisfied only in the region of the phase space enclosed by the inviscid solution (\ref{eq:xi}), that is the area inside the black curve in Fig. \ref{fig:phase}. Therefore, physically meaningful current profiles correspond to phase lines in this region.  
The curve $D(\xi, y)=0$ marks the critical condition and separates the upper reach of the phase lines corresponding to subcritical streams ($Fr<1$) from the lower one that refers to supercritical streams ($Fr>1$). 
{ It is interesting to note} that this critical divide does not depend on the friction parameter $\alpha$. In the case of $\alpha>0$ (i.e., diverging streams, flowing along increasing $\xi$), the phase space contains a focus where $N=D=0$ -- with coordinates $\{\xi_f=\sqrt[5]{2^4 / \alpha^3}, y_f=\sqrt[5]{\alpha^2 /2} \}$. The latter falls within the physically meaningful domain only if $\alpha<8/(9\sqrt{3})$.  

The transition from supercritical to subcritical streams takes place through the formation of a hydraulic jump. Typically, these appear as turbulent bores, although a series of standing waves (the so-called undular jump) can occur when the two streams before and after the jump are close to the critical condition (not considered here). The radial position of the hydraulic jump is obtained by applying the momentum principle along the flow direction to a short reach of circular sector of stream. This yields
\begin{equation}
\label{eq:mom_jump}
\frac{\gamma}{2}h^2 \phi r + \beta \rho (\phi r q) v=const.,     
\end{equation}
where the two addends refer to the hydrostatic and dynamical force, respectively, $\gamma$ is the fluid specific weight, $\phi$ is the central angle of the circular sector, $\beta$ is the momentum coefficient, and $\rho$ is the fluid density. In the previous relation, bed friction and lateral hydrostatic components have been neglected.

By introducing the dimensionless quantities and assuming turbulent motion ($\beta \simeq 1$), the previous relation (\ref{eq:mom_jump}) gives \begin{equation}
y^2+\frac{4}{\xi^2 y} \equiv F(\xi, y) = \mathrm{const.},    
\end{equation}
namely the (dimensionless) specific force $F$ that the two streams have to balance immediately before ($Fr>1$) and after ($Fr<1$) the hydraulic jump \cite{chow59}, i.e., $F(\xi_j,y_1)=F(\xi_j,y_2)$, where the subscripts 1 and 2 refer to the supercritical and subcritical depths at the radial position, $\xi_j$, of the hydraulic jump. A minimum of the specific force, $F=F_{min}$, occurs at $\xi=\sqrt[3]{2/\xi^2}$, in correspondence to the critical condition, $D=0$.

It is important to note that the hydraulic jump is a strongly dissipative phenomenon, related to the formation of turbulence and vorticity. In dimensionless terms, the energy dissipation of the hydraulic jump can be calculated as the difference in total energy across the jump, $[\Delta (\Delta+2y_1)/(y_1^2 y_2^2 \xi_j)-\Delta]$, where $\Delta=(y_2-y_1)$ is the depth difference across hydraulic jump.

\begin{figure*}
  \centering
  \includegraphics[width=150mm]{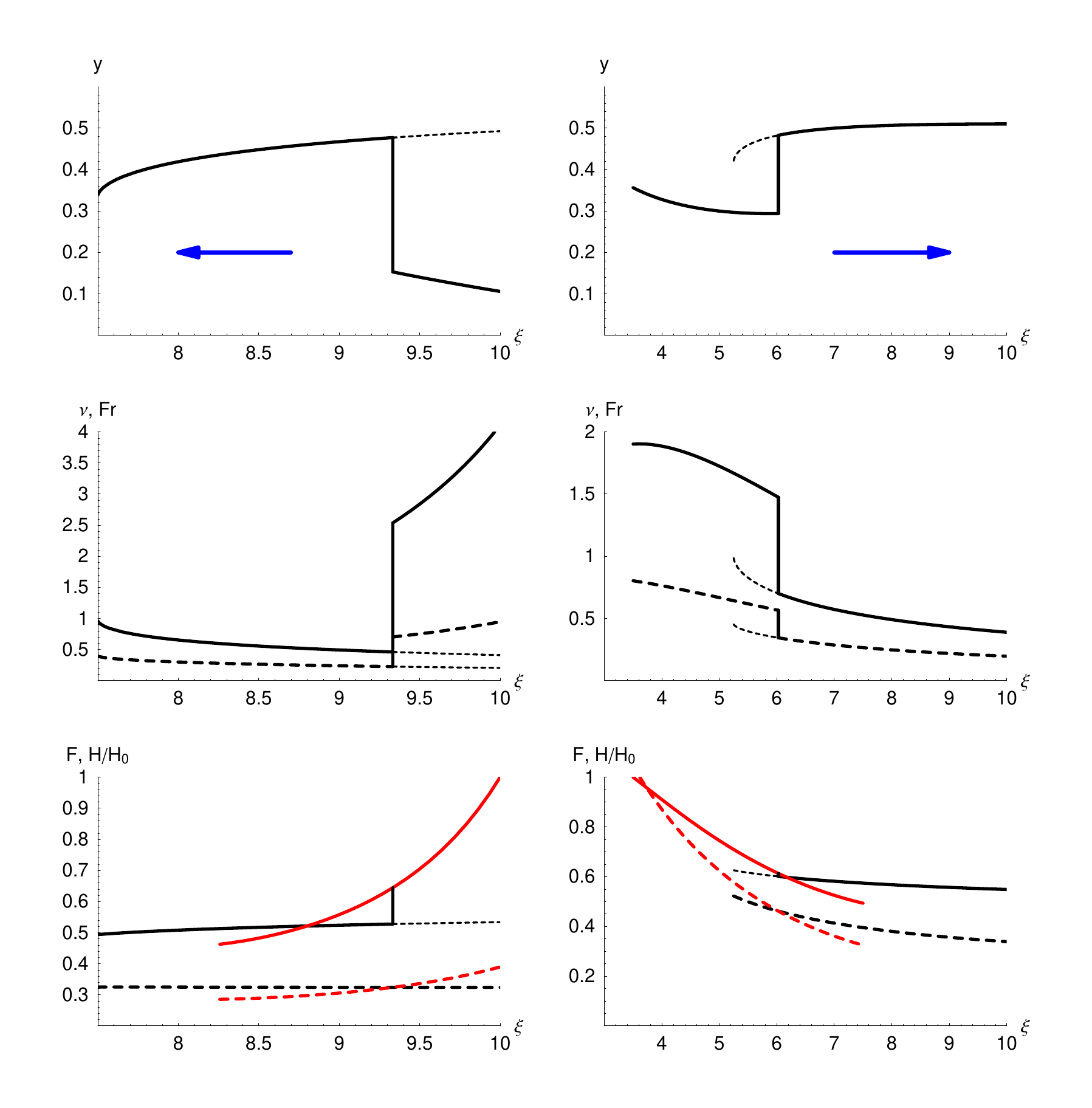}
  \caption{Dissipative solutions. The columns (see solid and dashed lines) refer to neutron star ($\alpha=-0.1$) and white hole ($\alpha=0.1$) cases, respectively. The first row shows stream profiles (solid lines; blue arrows indicate flow direction); the central row reports behaviors of velocity (dashed lines) and Froude number; the lower row displays behaviors of specific force (dashed lines) and stream head (referred to the initial stream energy, $H_0$). In the lower row, the red and black lines refer to the supercritical and subcritical streams, respectively. In all panels, dotted lines refer to profiles with no hydraulic jumps: the drain case (left column) and the spring case (right column).}
  \label{fig:profile_viscid}
\end{figure*}

In terms of astrophysical analogs, the introduction of friction in the shallow water dynamics 
{ brings about} additional cases, exemplified in Fig. \ref{fig:profile_viscid}: a neutron star (left column) and an analog of the dissipative white hole (right column); two other analogs are depicted in the same figure (with dotted lines) and correspond to dissipative forms of the black and white holes. In the case of the neutron-star analog, shown on the left of Fig.\ \ref{fig:profile_viscid}, the flows proceeds towards the center. Practically speaking, the fluid 
{ enters} the domain 
{ from} a { circular} sluice gate, placed along the external radius, and is drained through a central hole (of dimensions larger than $\xi_{min}$). The flow is initially supercritical and becomes subcritical after a hydraulic jump. The central and bottom panels show that the Froude number exhibits a non-monotonic behavior, as it first decreases in the supercritical reach and then increases when the stream becomes subcritical. The reason for this lies in the hydraulic constraint that the current must 
{ become critical} at the edge of the central hole, as shown in the right central panel, where the stream reaches $Fr=1$ on the hole edge placed at $\xi=7.5$. Accordingly, also the velocity shows a non-monotonic behaviour. Finally, the left lower panel highlights that the hydraulic jump entails an abrupt energy dissipation, which occurs where the specific forces of supercritical and subcritical streams are equal.

In the case of the dissipative white hole, illustrated in the right panels of Fig. 3, the fluid flows along increasing values of  $\xi$ (e.g., as if coming from a vertical jet impinging the bed close to $\xi = 0$) and { it is} initially in  supercritical conditions; a hydraulic jump then connects the profile to the subcritical one downstream. The central and bottom panels show that both the Froude number and the stream velocity decrease monotonically along the radius (although they would start increasing again, if the profile were to be continued), with a step change at the hydraulic jump. The condition of equality of the specific force dictates the radial position of the hydraulic jump, where a localized energy dissipation occurs. The subcritical profile $y(\xi)$ can be non-monotonic, since a maximum can occur depending on whether the subcritical reach intersects { or does not intersect} the 
{ line} $N=0$, { which is the} green line in Fig. \ref{fig:phase}.

Fig. \ref{fig:profile_viscid} reports also the profiles occurring when water drains or flows from a central hole { or spring}, without hydraulic jumps. Such profiles, corresponding to dissipative black holes, are characterized by subcritical streams shown as dotted lines. In the drain case (left panels), the flow originates from an external circular reservoir, then it accelerates converging towards the center and finally enters the hole in critical conditions. In the spring case (right panels), a profile, analogous to a dissipative white hole, starts from the critical condition ($Fr=1$, where water emerges), gradually slows down and joins the subcritical profile previously described in the case of white hole with shock. 

\begin{figure*}
  \centering
  \includegraphics[width=80mm]{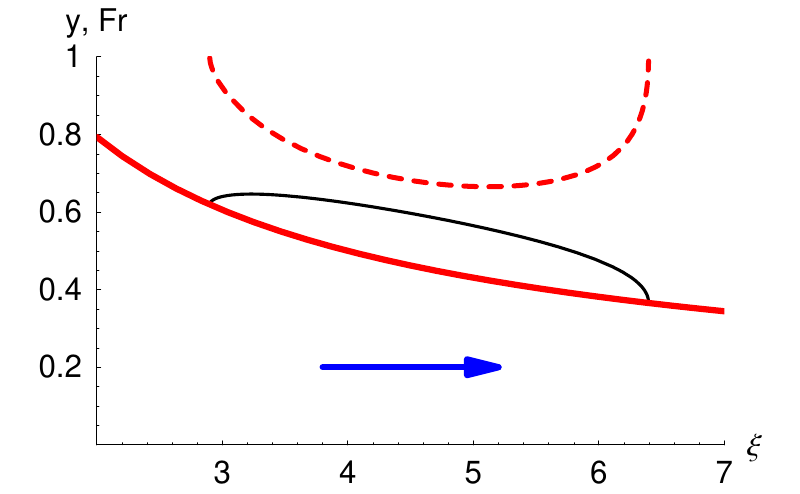}
  \caption{2-horizon spring: dissipative, subcritical profile connecting two horizons ($\alpha=0.4$, black line). Blue arrow indicates flow direction, while red lines refer to the critical conditions (solid line) and Froude number along the profile (dashed line), respectively. }
  \label{fig:profile_two horizon}
\end{figure*}

These last two subcritical analogs { with no hydraulic jumps} are similar to those discussed in the isentropic case. However, the presence of a maximum in the 
{ curve} $y=y(\xi)$ of the spring case, { which does not occur in isentropic situations, highlights}
an interesting class of profiles, connecting two critical horizons, that is peculiar to the viscous case. An example of a white hole confined between two horizons (see \cite{schutzhold2002gravity}, Sec. XI) is shown in Fig. \ref{fig:profile_two horizon}, where the flow springs from critical conditions, reaches a maximum and then decreases returning to the critical condition before jumping off from the outer edge of circular plate. 

The stability of the hydraulic jumps occurring in both analogs of Fig.\ \ref{fig:profile_viscid} is an interesting matter. If one only considers the specific forces $F$, they both appear spatially stable: perturbations of their radial position are absorbed by the consequent imbalance between the upstream and downstream specific forces, so that eventually the jumps return to their original position. However, a more detailed momentum balance across the jump, that included lateral hydrostatic pressures and bed friction, could alter this picture (see also \cite{ellegaard1998creating,ivanova2019structure}), especially in convergent cases \cite{foglizzo2012shallow}, as 
in one-dimensional streams in convergent or upward sloping channels, \cite{Marchi04,defina08}.  
Finally, it is worth mentioning that hydraulic jumps connecting supercritical to subcritical streams are possible also in the isentropic case. However, unlike the dissipative cases, their spatial position is undetermined, being marginally stable \cite{Valiani16}.  

\section{Conclusions}

The solutions of the shallow water equations present a variety of configurations, which besides their direct fluid dynamic interest may also have useful implications as analogs of specific astrophysical phenomena. For conditions of circular symmetry, the resulting steady state solutions have been discussed with particular attention to the transition between subcritical and supercritical conditions. The main cases are organized in Table I.
These steady state solutions may be realized in the laboratory and { can} be used as base solutions to explore the modes of propagation of disturbances and instabilities.

Starting from these configurations, several avenues for future research are suggested by the astrophysical analogies. Of particular interest is the stability of the hydraulic jumps. As already mentioned, this analysis is complicated by the presence of bottom friction and, in particular, by the pressure forces along the circumference of the shock, whose quantification depends on the specific geometry of the hydraulic jump \cite{Marchi04, defina08}. { Moreover, they} may include oscillation and symmetry breaking instabilities \cite{ellegaard1998creating,ivanova2019structure}, including those nicely documented in the neutron star analogue \cite{foglizzo2012shallow}.

Along a similar line, one could conjecture the appearance of roll waves, i.e.,\ pulsing and breaking waves, see \cite{dressler1949mathematical,whitham2011linear}), that could be realized in supercritical conditions with variable bottom slope. Apparently similar star-pulsation phenomena are well known in the literature \cite{balmforth1990effluent,balmforth1992solar, andersson1996gravitational}. In general, modifications of the bed slope (both downward and upward) introduce a degree of freedom, which would allow for the interplay between energy dissipation by friction and potential energy gain/loss to widen the gamut of hydraulic profiles and shock behaviors.   

Finally, including rotations  would be of interest for both Kerr-Newman black holes and for exploring wave generation in vorticity-shock interactions \cite{klein1994hydrodynamic,ellzey1995interaction}, while capillarity effects are known to generate lower wave-number disturbances in the upstream reach of obstacles \cite{whitham2011linear}, which have been linked to the Hawking radiation of black hole evaporation \cite{weinfurtner2011measurement}.
Extended thermodynamic formalism for turbulent flows, shocks and waves might also provide avenues to more concretely link black hole entropy to classical thermodynamics \cite{biro2020volume, porporato2020prsa}.

\begin{widetext}
\begin{center}
\begin{table}
\caption{List of shallow water analogs in circular symmetry  (SUB=Subcritical flow; SUP=Supercritical flow; HJ=Hydraulic Jump; SFH=Smooth Froude Horizon).} \begin{tabular}{ p{2.4cm}|p{2.1cm}|p{1.9cm}|p{2.5cm}|p{1.8cm}|p{2.3cm}|p{2.0cm}}
\hline
Shallow Water & Astr. Analog & Flow Dir. & Flow Types & Energetics & Eq./Fig. &Ref.\\
 \hline
Circular Jump & Neutron Star & Convergent & SUP$>$HJ$>$SUB & Dissipative & Eq. (\ref{eq:mom2D4}), Fig. 3 &\cite{foglizzo2012shallow}\\
Drain   & Turbulent Black Hole    & Convergent & SUB$>$SFH & Dissipative & Eq. (\ref{eq:mom2D4}), Fig. 3 &\\
Inviscid Drain & Black Hole    & Convergent  & SUB$>$SFH & Isentropic & Eq. (\ref{eq:xi}), Fig. 1 & \cite{schutzhold2002gravity}\\
Inviscid Spring  & White Hole & Divergent  & SUB$>$SFH & Isentropic & Eq. (\ref{eq:xi}), Fig. 1 & \cite{schutzhold2002gravity} \\
Spring & Turbulent White Hole & Divergent & SUB$>$SFH & Dissipative & Eq. (\ref{eq:mom2D4}), Fig. 3 &\\
2-Horizon Spring & Confined White Hole & Divergent & SFH$>$SUB$>$SFH & Dissipative & Eq. (\ref{eq:mom2D4}), Fig. 4 & \cite{schutzhold2002gravity}, Sect. XI \\
Circular Jump & White Hole with Shock & Divergent  & SUP$>$HJ$>$SUB & Dissipative & Eq. (\ref{eq:mom2D4}), Fig. 3 & \cite{Bhattacharjee2016,Volovik2005}\\
 \hline
\end{tabular}
\end{table}
\end{center}
\end{widetext}

{\it Acknowledgements. --}  A.P. the US National Science Foundation (NSF) grants EAR-1331846 and EAR-1338694. LR acknowledges that the present research has been partially supported by MIUR grant Dipartimenti di Eccellenza 2018-2022 (E11G18000350001).



\providecommand{\noopsort}[1]{}\providecommand{\singleletter}[1]{#1}%
\end{document}